\documentclass[twocolumn]{aastex62}

\newcommand{\planck}{\textsl{Planck}}

\newcommand{\lcdm}{\ensuremath{\Lambda\mathrm{CDM}}}

\usepackage{bm} 
\usepackage{mathtools}

\graphicspath{{./}{figures/}}

\shorttitle{Correlated Data Sets}
\shortauthors{Kable et al.}

\begin{document}

\title{Analytic Calculation of Covariance between Cosmological Parameters from Correlated Data Sets, with an Application to SPTpol}

\correspondingauthor{Joshua A. Kable}
\email{jkable2@jhu.edu}

\author[0000-0002-0786-7307]{Joshua A. Kable}
\affil{Johns Hopkins University \\
3400 North Charles Street \\
Baltimore, MD 21218, USA}

\author{Graeme E. Addison}
\affiliation{Johns Hopkins University \\
3400 North Charles Street \\
Baltimore, MD 21218, USA}

\author{Charles L. Bennett}
\affiliation{Johns Hopkins University \\
3400 North Charles Street \\
Baltimore, MD 21218, USA}

\begin{abstract}
Consistency checks of cosmological data sets are an important tool because they may suggest systematic errors or the type of modifications to \lcdm\ necessary to resolve current tensions. In this work, we derive an analytic method for calculating the level of correlations between model parameters from two correlated cosmological data sets, which complements more computationally expensive simulations.  This method is an extension of the Fisher analysis that assumes a Gaussian likelihood and a known data covariance matrix. We apply this method to the SPTpol temperature and polarization CMB spectra (TE and EE).  We find weak correlations between \lcdm\ parameters with a 9$\%$ correlation between the TE-only and EE-only constraints on $H_0$ and a 25$\%$ and 32$\%$ correlation for log($A_s$) and $n_s$ respectively. The TE-EE parameter differences are consistent with zero, with a PTE of 0.53. Using simulations we show that this test is independent of the consistency of the SPTpol TE and EE bandpowers with the best-fit \lcdm\ model spectra. Despite the negative correlations between the TE and EE  power spectra, the correlations between TE-only and EE-only \lcdm\ parameters are positive. Ignoring correlations in the TT-TE and TE-EE comparisons biases the $\chi^2$ low, artificially making parameters look more consistent. Therefore, we conclude that these correlations need to be accounted for when performing internal consistency checks of the TT vs TE vs EE power spectra for future CMB analyses.

\end{abstract}


\section{Introduction}
Cosmic Microwave Background (CMB) experiments show good agreement with \lcdm\ \citep[e.g.,][]{bennett/etal:2013,planck/6:2018,sievers/etal:2013,story/etal:2013}, with mild tension reported by SPTpol \citep{Henning/etal:2018}. However, stronger tensions emerge when CMB experiments are compared to some other cosmological experiments. The most notable tension is in the determination of the present expansion rate of the universe or Hubble constant, $H_0$. There is currently a $4.4\sigma$ tension between the most recent \planck\ results \citep{planck/6:2018} and the cosmological distance ladder measurement \citep{Riess/etal:2019}. When the distance ladder is combined with strong gravitational lensing time delays by the $H_0$ Lenses in COSMOGRAIL's Wellspring (H0LiCOW), the discrepancy with \planck\ is 5.3$\sigma$ \citep{Holicow/etal:2019}. \cite{freedman/etal:2019} used the Tip of the Red Giant Branch as a calibration for Type Ia supernovae and find only a 1.2$\sigma$ difference from \planck. However, \cite{Yuan/etal:2019} argue that \cite{freedman/etal:2019} overestimate the Large Magellanic Cloud extinction. Accounting for this, their result is in 2.5$\sigma$ tension with \planck\ while consistent at 0.7$\sigma$ with \cite{Riess/etal:2019}.

While \planck\ is the most precise CMB experiment to date, the Hubble tension persists when Baryon Acoustic Oscillations (BAO) data are combined with other CMB experiments or even deuterium abundances \citep[e.g.,][]{addison/etal:2018}. Additionally, \planck\ prefers a 2-3$\sigma$ larger value of $S_8 = \sigma_8 \sqrt{\frac{\Omega_m}{0.3}}$, which is a measure of matter clustering, than weak lensing surveys \citep[e.g.,][]{planck/6:2018,Hildebrandt/etal:2018,Joudai/etal:2018,Abbott/etal:2018,Hikage/etal:2019} and cluster abundance surveys \citep[e.g.,][]{Lin/etal:2017,McCarthy/etal:2018}.

In the last five years, these tensions have sparked a keen interest in assessing the consistency of cosmological data sets. For the $H_0$ tension, because any one cosmological experiment can be removed and not eliminate the discordance, it is unlikely that the discordance is the result of underestimated or unmodeled systematics. Nevertheless, these consistency tests may help illuminate what extensions to \lcdm\ the data prefer. 

There are essentially two ways to assess the consistency of cosmological data sets: (1) comparing the data directly or (2) comparing the resulting parameter constraints. For CMB experiments, directly comparing the data can be done at either the map level \citep[e.g.,][]{louis/etal:2014,larson/etal:2015,Hou/etal:2018} or at the power spectrum level \citep[e.g.,][]{Hou/etal:2018,Huang/etal:2018,Mocanu/etal:2019}.  Changing the multipole moments included in the fit impacts the best-fit cosmology, and is a valuable internal consistency check \citep{addison/etal:2016,planck/51:2017,Aylor/etal:2017}. Additionally, there have been investigations into the correlations or degeneracies between parameters expected in \lcdm\ or possible extensions \citep[e.g.,][]{Kable/etal:2019,Huang/etal:2019}.

In this paper, we provide a method for determining whether parameter constraints from correlated data are consistent. These correlations between parameters arise when data sets or sub-sets share correlated noise or sample variance. We derive an analytic method for calculating the covariance between parameters from correlated data sets or sub-sets using a known data covariance matrix.  This is an extension of the traditional Fisher analysis \citep[e.g.,][]{Heavens:2009,Verde:2010}, providing a fast alternative to more computationally intensive simulations \citep[e.g.,][]{planck/51:2017,Sanchez/etal:2017,Louis/etal:2019}.

We apply this analytic method to CMB temperature and polarization constraints. Looking at the consistency of temperature and polarization is valuable now that polarization data provide comparable constraining power on \lcdm\ parameters \citep{Galli/etal:2014, planck/6:2018,Henning/etal:2018}. Further, we conclude that this analytic method is applicable to many cosmological data sets.

This paper is organized as follows. In Section 2, we derive our analytic method for calculating the covariance between parameters from correlated data. In Section 3, we apply this to SPTpol TE and EE power spectra to determine the consistency of the parameter constraints. In Section 4, we discuss the correlations and consistency between TT, TE, and EE CMB power spectra for a cosmic variance limited experiment. Finally, in Section 5, we provide conclusions. 

\section{Derivation of Analytic Method}
The goal of this section is to quantify the correlation between parameters constrained by two correlated data sets or sub-sets with known data covariance. We derive an expression for how the maximum likelihood parameters deviate from a set of fiducial parameters in terms of the difference between the data and the fiducial mean vector. This allows us to compute the covariance between the parameters from correlated data sub-sets in terms of quantities that are known or easy to calculate: the data covariance and derivatives of the mean vectors with respect to parameters. 

We make the same basic assumptions as Section 2 of \cite{Huang/etal:2019}, namely: 
\begin{itemize}
    \item We assume a Gaussian likelihood with a data covariance that is independent of the cosmological parameters, with parameter dependence only entering through the mean values. 
    \item We assume that the data are sufficiently constraining that they may be treated as a linear perturbation about a fiducial theory model. Similarly, the maximum likelihood parameters are treated as a linear perturbation about the fiducial parameters, for each data sub-set.
\end{itemize}

This is equivalent to saying that the maximum likelihood parameters are Gaussian distributed and that the Fisher matrix accurately describes the covariance of the maximum likelihood parameters. Assuming diffuse priors, it follows that the Bayesian posterior parameter distribution is Gaussian with the same covariance as the maximum likelihood parameters. More discussion can be found in \cite{Raveri/etal:2019}.

In this case, the log-likelihood can be defined up to a constant as 
\begin{equation}
    \mathrm{log}(\mathcal{L}) = -\frac{1}{2}(\textbf{d} - \bm{\mu}(\bm{\theta}))^T\pmb{\mathbb{C}}^{-1}(\textbf{d} - \bm{\mu}(\bm{\theta}))
\end{equation}
where $\textbf{d}$ is the measured data vector, $\bm{\mu}(\bm{\theta})$ is the theory data vector given for some parameter vector $\bm{\theta}$, and $\pmb{\mathbb{C}}$ is the data covariance matrix. The derivative of the log-likelihood with respect to the parameters evaluated at the maximum likelihood parameter vector is defined to be zero. Namely,
\begin{equation}
    \bm{0} = \frac{\partial \mathrm{log}(\mathcal{L})}{\partial \bm{\theta}} \Bigg|_{\textrm{ML}} = \frac{\partial \bm{\mu}^T}{\partial \bm{\theta}}\Bigg|_{\textrm{ML}}\pmb{\mathbb{C}}^{-1}(\textbf{d} - \bm{\mu}(\bm{\theta_{\textrm{ML}}}))
\end{equation}
where $\frac{\partial \bm{\mu}^T}{\partial \bm{\theta}}\big|_{\textrm{ML}}$ is a matrix with elements $\frac{\partial \mu_i}{\partial \theta_j}$. Assuming the parameter vector is near the fiducial parameter vector, $\bm{\theta} \approx \bm{\theta}_{\textrm{fid}}$, the theory vector can be Taylor expanded to linear order so that 

\begin{equation}
    \frac{\partial \mathrm{log}(\mathcal{L})}{\partial \bm{\theta}}\Bigg|_{\textrm{ML}} = \frac{\partial \bm{\mu}^T}{\partial \bm{\theta}}\Bigg|_{\textrm{fid}}\pmb{\mathbb{C}}^{-1}\left(\bm{\delta} -\frac{\partial \bm{\mu}}{\partial \bm{\theta}}\Bigg|_{\textrm{fid}}(\bm{\theta}_{\textrm{ML}} - \bm{\theta}_{\textrm{fid}})\right)
\end{equation}
where $\bm{\delta} = \textbf{d} - \bm{\mu}(\bm{\theta_{\textrm{fid}}})$.

Note that the derivative of the theory vector is now evaluated at the fiducial parameter values instead of the maximum likelihood parameter values. This is true to linear order with the assumption that $\bm{\delta}$ is small. This implies that 
\begin{equation}
    \bm{0} = \frac{\partial \bm{\mu}^T}{\partial \bm{\theta}}\Bigg|_{\textrm{fid}}\pmb{\mathbb{C}}^{-1}\bm{\delta} - \bm{F}(\bm{\theta}_{\textrm{ML}} - \bm{\theta}_{\textrm{fid}})
\end{equation}
where $\bm{F}$ is the parameter Fisher matrix 
\begin{equation}
    \bm{F} = \frac{\partial \bm{\mu}^T}{\partial \bm{\theta}}\Bigg|_{\textrm{fid}}\pmb{\mathbb{C}}^{-1}\frac{\partial \bm{\mu}}{\partial \bm{\theta}}\Bigg|_{\textrm{fid}}.
\end{equation}
The Fisher matrix is the inverse of the parameter covariance matrix. Rearranging Equation (4) gives 
\begin{equation}
    \bm{\theta}_{\textrm{ML}} = \bm{\theta}_{\textrm{fid}} + \bm{F}^{-1}\frac{\partial \bm{\mu}^T}{\partial \bm{\theta}}\Bigg|_{\textrm{fid}}\pmb{\mathbb{C}}^{-1}\bm{\delta}.
\end{equation}
Taking the expectation value over many realizations of the data vector $\bm{d}$, $\langle \bm{\theta}_{\textrm{ML}}\rangle=\langle \bm{\theta}_{\textrm{fid}} \rangle$ as $\langle \bm{\delta} \rangle =\langle \textbf{d} - \bm{\mu}(\bm{\theta_{\textrm{fid}}}) \rangle=0$. The covariance between maximum likelihood parameters from a data set X and the maximum likelihood parameters from a data set Y is determined to be
\begin{equation}
    \langle(\bm{\theta}_{\textrm{ML}}^X - \langle \bm{\theta}_{\textrm{ML}}^X\rangle)(\bm{\theta}_{\textrm{ML}}^Y - \langle \bm{\theta}_{\textrm{ML}}^Y\rangle)\rangle = (\pmb{M}^{X})^{T}\pmb{\mathbb{C}}^{XY}\pmb{M}^{Y}
\end{equation}
where 
\begin{equation}
   \pmb{M}^{X} = (\pmb{\mathbb{C}}^{XX})^{-1}\frac{\partial \bm{\mu}^X}{\partial \bm{\theta}^X}\Bigg|_{\textrm{fid}}(\bm{F}^{XX})^{-1}
\end{equation}
and $\pmb{\mathbb{C}}^{XY} = \langle(\textbf{d}^X - \bm{\mu}^X)(\textbf{d}^Y - \bm{\mu}^Y)^T\rangle$ is the block of the data covariance matrix that describes the covariance between X and Y. When X and Y are the same, Equation 7 reduces to the inverse of the Fisher matrix. When X and Y are different Equation 7 maps the data covariance $\pmb{\mathbb{C}}^{XY}$ to the parameter covariance. While we derived Equation 7 in terms of maximum likelihood parameters, it also describes the covariance of the Bayesian posterior, provided the assumptions at the beginning of this section hold. See for example Chapter 4 and Appendix~B of \cite{gelman2013bayesian}. 
 
Note that the covariance of parameters from two correlated data sets cannot be calculated using a single Fisher matrix containing two sets of varying parameters (one set for each data set). This introduces additional coupling between X and Y and means that, for example, the XX and YY parameter covariance blocks do not correctly reduce to the inverse of the X-only or Y-only Fisher matrices. 

 \section{SPTpol: A Worked Example}
In the previous section we derived an analytic expression for the covariance matrix for parameters from two correlated data sets. In this section, we apply this formula to CMB polarization data.  \cite{Galli/etal:2014} showed that constraints from the TE power spectrum could be more constraining than constraints from either the TT or EE power spectra individually. While there is often a focus on the \planck\ TT power spectrum, constraints from the \planck\ TE power spectra are just as constraining for several of the parameters \citep{planck/6:2018}. The future improvement in \lcdm\ parameter constraints from the CMB will be from studies of the TE and EE power spectra. SPTpol provides some of the tightest constraints to date for the TE and EE power spectra \citep{Henning/etal:2018}.

The correlations between the TT, TE, and EE power spectra come from noise and cosmic variance. The multipole covariance matrix for a cosmic variance limited CMB experiment is given by
\begin{equation}
   \pmb{\mathbb{C}}_{\ell} =  
  \left[ {\begin{array}{ccc}
  (\pmb{\mathbb{C}}_{\ell})_{TTTT} & (\pmb{\mathbb{C}}_{\ell})_{TTTE} & (\pmb{\mathbb{C}}_{\ell})_{TTEE} \\
   (\pmb{\mathbb{C}}_{\ell})_{TTTE} & (\pmb{\mathbb{C}}_{\ell})_{TETE} & (\pmb{\mathbb{C}}_{\ell})_{TEEE} \\
   (\pmb{\mathbb{C}}_{\ell})_{TTEE} & (\pmb{\mathbb{C}}_{\ell})_{TEEE} & (\pmb{\mathbb{C}}_{\ell})_{EEEE} \\
  \end{array} } \right].
\end{equation}

The covariance sub-blocks for the TT, TE, and EE power spectra are approximately given by

\begin{equation}
\begin{aligned}
    (\pmb{\mathbb{C}}_{\ell})_{TTTT} &= \frac{2}{(2\ell + 1)f_{sky}}(C_{\ell}^{TT})^2 \\
    (\pmb{\mathbb{C}}_{\ell})_{TTTE} &= \frac{2}{(2\ell + 1)f_{sky}}(C_{\ell}^{TT}C_{\ell}^{TE}) \\
   (\pmb{\mathbb{C}}_{\ell})_{TTEE} &= \frac{2}{(2\ell + 1)f_{sky}}(C_{\ell}^{TE})^2 \\
    (\pmb{\mathbb{C}}_{\ell})_{TETE} &= \frac{1}{(2\ell + 1)f_{sky}}(C_{\ell}^{TT}C_{\ell}^{EE} + (C_{\ell}^{TE})^2) \\
    (\pmb{\mathbb{C}}_{\ell})_{TEEE} &= \frac{2}{(2\ell + 1)f_{sky}}(C_{\ell}^{TE}C_{\ell}^{EE}) \\ 
    (\pmb{\mathbb{C}}_{\ell})_{EEEE} &= \frac{2}{(2\ell + 1)f_{sky}}(C_{\ell}^{EE})^2,\\
\end{aligned}
\end{equation}
where $f_{sky}$ is the fraction of the sky observed (e.g., \citealt{scott/srednicki/white:1994}; Section~2.6 of \citealt{weinberg2008cosmology}). It is because of the correlation between the CMB power spectra that there is a correlation between the CMB parameter constraints from each of the spectra.

The covariance at high multipoles includes non-Gaussian lensing terms \citep[e.g.,][]{benoit-levy/etal:2012,Manzotti/etal:2014,Motloch/etal:2019}. We do not include these effects in Section 4 where we investigate the correlations for the cosmic variance limited case for a wide range of multipole moments. Since collaborations will have to include these effects in their band-power covariance matrices, it will not be extra work to include them when calculating the parameter covariance matrix between different data sub-sets. 
 
\subsection{SPTpol}

The South Pole Telescope Polarimeter (SPTpol) measures 500 square degrees of the southern hemisphere sky. \cite{Henning/etal:2018} report power spectra at 150 GHz taken over three observing seasons. 

The SPTpol collaboration provides binned TE and EE power spectra, a binning matrix, and a band-power covariance matrix \footnote{https://pole.uchicago.edu/public/data/henning17/}. In Figure 1, we compare the correlation between TE and EE power spectra from the band-power covariance matrix for SPTpol to another contemporary high resolution CMB polarization experiment, Atacama Cosmology Telescope Polarimeter (ACTPol) \citep{louis/etal:2017}, and to the cosmic variance limited case. Adding noise to the band-power covariance matrix results in weaker correlations between the TE and EE power spectra. Both experiments follow the cosmic variance limited case at low multipoles, but as the covariance becomes dominated by noise, the correlations tend to zero. \cite{Henning/etal:2018} report that the SPTpol TE spectrum is sample variance limited at $\ell < 2050$ and the EE spectrum is sample variance limited at $\ell < 1750$.

The correlation between the TE and EE spectra is oscillatory about zero for low multipole moments, but it is negative for high multipole moments making the two data sets predominantly negatively correlated. The negative correlation is a result of the mostly negative values of the TE power spectrum, $C_{\ell}^{TE}$, at high multipole moments because the correlation for the cosmic variance limited case is given by 
\begin{equation}
    \frac{(\pmb{\mathbb{C}}_{\ell})_{TEEE}}{\sqrt{(\pmb{\mathbb{C}}_{\ell})_{TETE}(\pmb{\mathbb{C}}_{\ell})_{EEEE}}} = \frac{C_{\ell}^{TE}}{\sqrt{C_{\ell}^{TT}C_{\ell}^{EE}+(C_{\ell}^{TE})^2}}
\end{equation}
because $C_{\ell}^{EE}$ is necessarily positive. 

SPTpol compared their measured TE and EE power spectra to the \lcdm\ predicted power spectra using the maximum likelihood parameters from the fit to the data (i.e. comparing the data vector $\textbf{d}$ to the theory vector $\bm{\mu}$). For the joint TE + EE fit, they report a probability to exceed (PTE) of 0.017, indicating a poor fit. For TE-only and EE-only they report a PTE of 0.045 and 0.12, respectively. They conclude that it is difficult to assess the consistency of the different \lcdm\ solutions because the parameter constraints exhibit varying degrees of degeneracy. Motivated by this, we investigate the parameter consistency of the SPTpol TE-only and EE-only mean parameter vectors. 

We ran three Markov Chain Monte Carlo (MCMC) samplers on the SPTpol likelihoods including a TE-only, EE-only, and joint fit using the \texttt{CosmoMC}\footnote{\url{https://cosmologist.info/cosmomc/}} package \citep{lewis/bridle:2002}. We summarize the mean and 68$\%$ credible intervals for a set of fit and derived \lcdm\ parameters in Table 1. The first five parameters are fit, while the remainder are derived based on fit parameters. The parameter $\theta_{MC}$ is an approximation of the ratio of the sound horizon to the angular diameter distance to the surface of last scattering that \texttt{CosmoMC} fits instead of $\theta_*$. For definitions of each of these parameters see Table 1 of \cite{planck/16:2013}. We fix $\tau = 0.078$ because the SPTpol TE and EE power spectra do not add much information and this is the central value that SPTpol used for their prior. Note that this artificially tightens the $A_s$ uncertainty. We use the same priors on the SPTpol nuisance parameters as \cite{Henning/etal:2018}. We use the joint fit cosmology in Table 1 throughout this work as our fiducial model. 

\begin{deluxetable}{cccc}
\tabletypesize{\footnotesize}
\tablewidth{3.5in}

 \tablecaption{Mean values and 68$\%$ credible intervals for \lcdm\ parameters for the MCMC chains for the SPTpol joint TE and EE fit, the SPTpol TE-only fit, and the SPTpol EE-only fit. In all cases we fixed the value of $\tau$ to the central value adopted by SPTpol (see text). \label{tab:best-fit}}

 \tablehead{
 \colhead{Parameter} & \colhead{Joint (TE + EE)} & \colhead{TE-only} & \colhead{EE-only} 
 }
 
 \startdata
 $\Omega_ch^2$ & 0.1091 $\pm$ 0.0046 & 0.1186 $\pm$ 0.0071 & 0.1058 $\pm$ 0.0072 \\
 $\Omega_bh^2$ &0.02296 $\pm$ 0.00046 & 0.02336 $\pm$ 0.00072 & 0.02235 $\pm$ 0.0011 \\
 $\log(A_s)$ & 3.025 $\pm$ 0.020 & 3.078 $\pm$ 0.037 & 3.000 $\pm$ 0.029 \\
 $n_s$ &0.998 $\pm$ 0.022 & 0.965 $\pm$ 0.030 & 1.032 $\pm$ 0.037 \\
 $100\theta_{MC}$ & 1.03992 $\pm$ 0.00082 & 1.0396 $\pm$ 0.0012 & 1.0408 $\pm$ 0.0012 \\
 \hline
 $100\theta_*$ & 1.04006 $\pm$ 0.00083 & 1.0396 $\pm$ 0.0013 & 1.0410 $\pm$ 0.0012 \\
 $H_0$ &71.6 $\pm$ 2.0 & 68.3 $\pm$ 2.6 & 72.8 $\pm$ 3.4 \\
 $\Omega_m$ & 0.260 $\pm$ 0.023 & 0.308 $\pm$ 0.039 & 0.246 $\pm$ 0.036 \\
 $\sigma_8$ & 0.770 $\pm$ 0.021 & 0.813 $\pm$ 0.032 & 0.760 $\pm$ 0.035 \\
 \enddata

 \vspace{-0.5cm}

\end{deluxetable}

\begin{figure}[!tbp]
  \begin{minipage}[b]{0.4\textwidth}
    \hspace*{-0.75cm}
    \includegraphics[width=4in, height=3in]{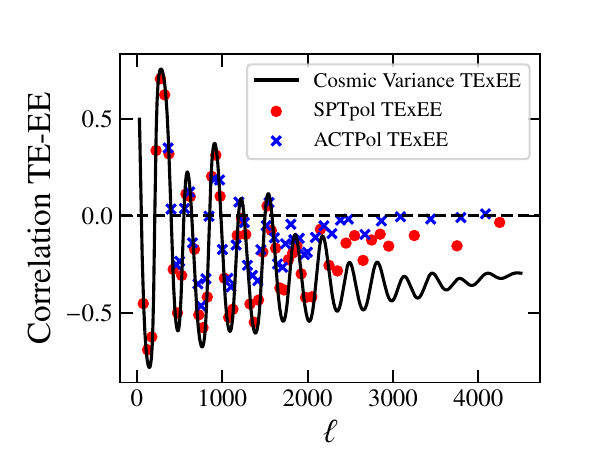}
  \end{minipage}
  \caption{ The correlation coefficient in the TE-EE band-power covariance matrices for SPTpol, ACTPol, and the cosmic variance limited case. For SPTpol and ACTPol, the correlations fall to zero at high $\ell$ because of noise. The correlation between the TE and EE data is oscillatory but predominantly negative. 
 \label{fig:l}}
  \end{figure}
  
\subsection{Analytic Solution for SPTpol Maximum Likelihood}
In this section, we test the validity of the analytic solution derived in Section 2. We do this by comparing the covariance matrix we calculate using the analytic solution to the covariance matrix we get from maximum likelihood simulations. For our test, we use SPTpol TE and EE power spectra. For the simulations, we
\begin{itemize}
    \item Calculate the fiducial TE and EE theory spectra and bin using the SPTpol binning matrices. We calculate these power spectra using $\texttt{Pycamb}$, the python wrapper for $\texttt{CAMB}$\footnote{https://camb.info/} \citep{lewis/etal:2000}, and the joint-fit mean cosmology shown in Table 1. 
    \item Generate 1000 sets of simulated SPTpol band-powers using the fiducial binned spectrum as mean and SPTpol band-power covariance matrices drawing from a multivariate Gaussian distribution. 
    \item Run the Maximum Likelihood finding algorithm in $\texttt{CosmoMC}$ for each case.
    \item Show the mean and sample covariance from the distribution of 1000 best-fit TE-only and EE-only parameter vectors as red contours in Figure 2. 
\end{itemize}

To calculate the covariance matrix using the analytic solution in Equation 7, we need to calculate the derivatives of the TE and EE power spectra with respect to the parameters. To calculate the derivative matrices, we use a finite difference method. We calculate TE and EE power spectra using $\texttt{Pycamb}$ and the parameters $\{\theta_*,\Omega_ch^2,\Omega_bh^2,\log(A_s),n_s\}$. We choose to use the joint fit mean values given in Table 1 as the fiducial cosmology. Because we have control over the fiducial cosmology for both the simulations and the analytic calculation, we defer further discussion of the effect of the choice of fiducial cosmology to the next section. 

We generate power spectra with multipoles $50 \leq \ell \leq 4500$ for the finite difference calculations of the derivative matrix to match SPTpol. For the numerical derivative, we use a step size of 1$\%$ but find that there is a negligible change in the derivatives resulting from changes in the step size from 0.5$\%$ to 2$\%$ indicating numerical stability of the derivative. 

Using the band-power covariance matrices and the TE and EE derivative matrices, we calculate the TE and EE parameter Fisher matrices using Equation 5. We calculate the covariance between $\bm{\theta}^{TE}_{\textrm{ML}}$ and $\bm{\theta}^{EE}_{\textrm{ML}}$ using Equation 7. This is a covariance matrix of 10 parameters which includes TE-only and EE-only versions of each of $\{\theta_*,\Omega_ch^2,\Omega_bh^2,\log(A_s),n_s\}$. In Figure 2, we show the parameter contours for a normal distribution centered at the joint fit mean cosmology with this covariance matrix in black. 

\begin{figure*}[!tbp]
  \begin{minipage}[b]{0.4\textwidth}
    \hspace*{0cm}
    \includegraphics[width=6.5in, height=6.5in]{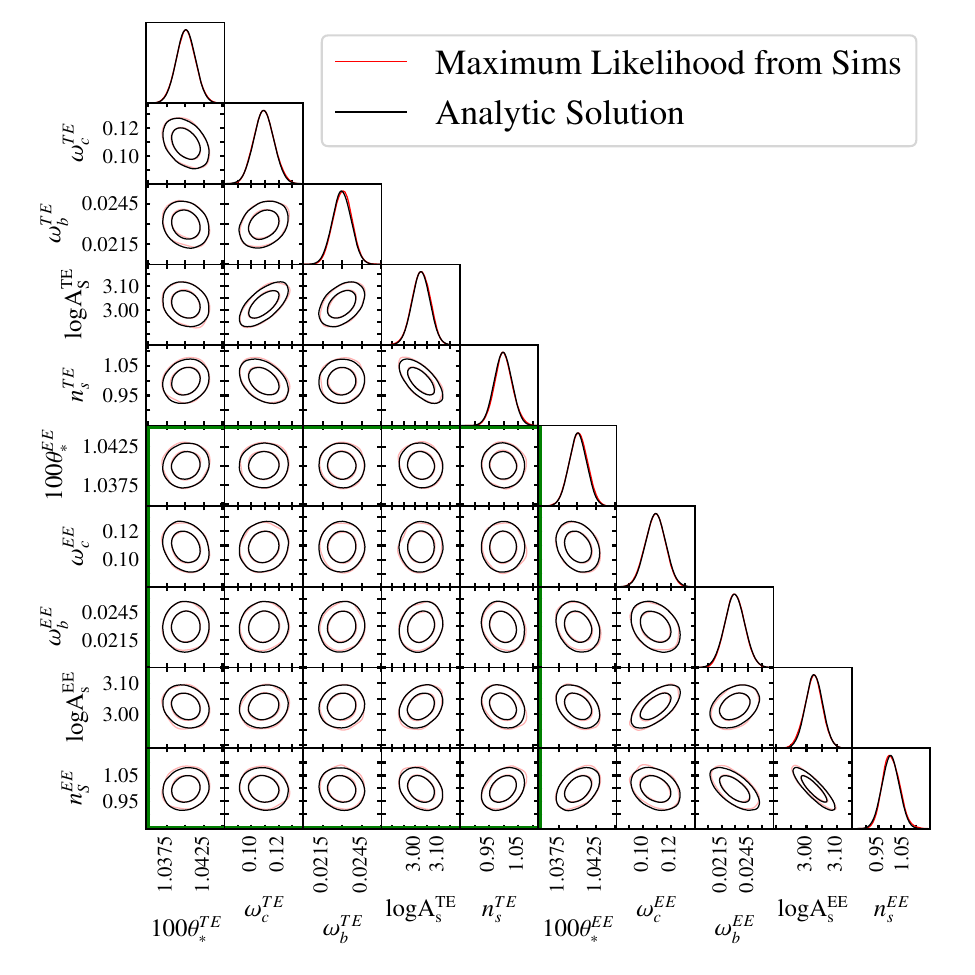}
  \end{minipage}
  \caption{Maximum likelihood parameter vectors from 1000 simulated SPTpol TE and EE power spectra (red),
    with the predicted distribution from the covariance matrix calculated using the analytic solution 
    given by Equation 7 (black). The contours are at 1 and 2 $\sigma$, and we use $\omega_x$ in place of $\Omega_xh^2$. The contours within the green box show the covariance between TE-only and EE-only parameters. Most of the TE-only parameters are correlated with their counterparts from the EE-only fit at around the 10$\%$ level; however, $\log(A_s)$ and $n_s$ have about $30\%$ correlation between TE-only and EE-only.
 \label{fig:2}}
  \end{figure*}

In Figure 2, the upper left and bottom right blocks correspond to the TE-only and EE-only parameter constraints, respectively. The consistency between the maximum likelihood simulations and the analytic solution in these blocks illustrates that the TE-only and EE-only Fisher matrix approach reproduces the constraining power of the data. The correlation between the TE-only and EE-only parameters are shown in the five by five block in the bottom left portion of Figure 2. The correlations between TE-only and EE-only parameters are weak: between 10$\%$ and 30$\%$. There is excellent consistency between the maximum likelihood (red) contours and the analytic solution (black) contours, which shows that the analytic solution achieves results comparable to the maximum likelihood simulation method. 

\subsection{Comparing SPTpol Data}
In the previous subsection, we showed that the analytic solution succeeds for simulated data. In this subsection, we use the analytic solution to compare the TE-only mean cosmology to the EE-only mean cosmology from the SPTpol data to determine whether they are consistent. For the TE-only and EE-only mean cosmologies we use the corresponding mean cosmologies from Table 1. For our five parameter vector, we choose to use $H_0$ instead of $\theta_*$ as it is more easily related to other cosmological experiments. We find that this choice makes a negligible difference to either the correlations or eventual determinations of consistency. 

We recalculate the covariance matrix between the TE-only and EE-only parameters again using the mean cosmology from the joint fit as the fiducial model and with a multipole range of $50 \leq \ell \leq 4500$. The constraints on the SPTpol nuisance parameters are largely driven by their priors, meaning their contribution to the cosmological parameter covariance matrix is negligible. To match the SPTpol data, we take into account the SPTpol aberration in the analytic calculation. The aberration is due to our motion relative to the CMB rest frame and causes a small shift in the measured power spectrum, which biases $\theta_{\textrm{MC}}$. SPTpol accounts for this by applying 
\begin{equation}
    C_{\ell} \rightarrow C_{\ell}\left(1 - \frac{d\log (C_{\ell})}{d\log (\ell)}\beta \langle \cos\theta\rangle\right),
\end{equation}
where $\beta = 1.23 \times 10^{-3}$ and $\langle \cos\theta \rangle = -0.4$. This aberration is applied to the theory power spectrum and not to the data. 

To compare the TE-only mean cosmology with the EE-only mean cosmology, we calculate the covariance matrix of the difference parameters given by
\begin{equation}
  \langle \bm{\theta}_{\mathrm{diff}}\bm{\theta}_{\mathrm{diff}}^T \rangle =  \pmb{\mathbb{C}}^{TE-TE} + \pmb{\mathbb{C}}^{EE-EE} - \pmb{\mathbb{C}}^{TE-EE} - \pmb{\mathbb{C}}^{EE-TE}
\end{equation}
where $\bm{\theta}_{\mathrm{diff}} = \bm{\theta}^{TE} - \bm{\theta}^{EE}$ and $\pmb{\mathbb{C}}^{X-Y}$ is the parameter covariance matrix between data sets X and Y. For our purposes, these are just the four sub-blocks of the covariance matrix we calculate using the analytic solution. Note that in the case of no correlation between data sets X and Y, $\pmb{\mathbb{C}}^{TE-EE} = \pmb{\mathbb{C}}^{EE-TE} = 0$.

\begin{figure}[!tbp]
  \begin{minipage}[b]{0.4\textwidth}
    \hspace*{-1cm}
    \includegraphics[width=4in, height=4in]{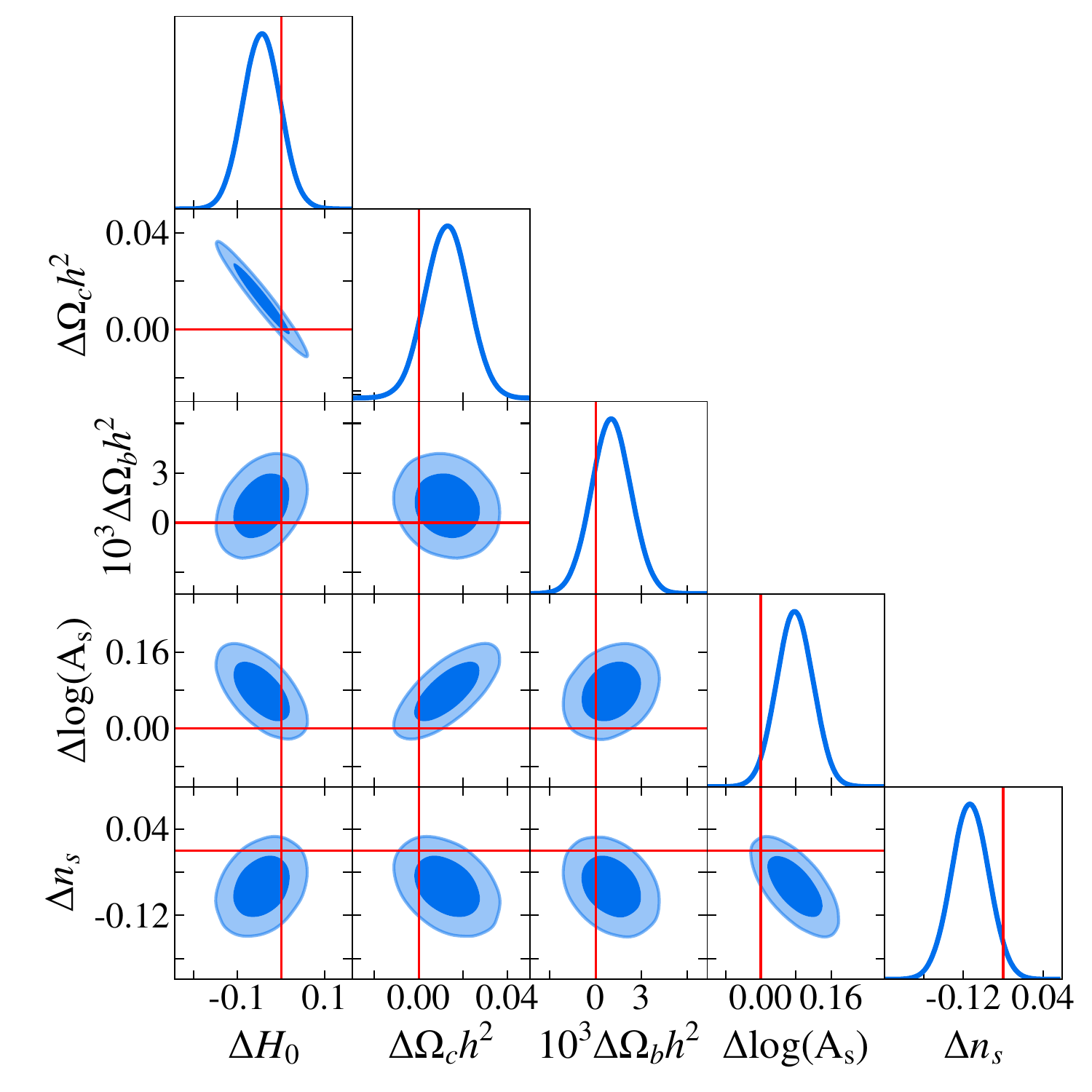}
  \end{minipage}
  \caption{Parameter contours for the difference of TE-only parameters and EE-only parameters. The red horizontal and vertical lines indicate zero, which is the expected value for the differences. The TE-only and EE-only parameter differences are consistent with zero with a PTE of 0.53. 
 \label{fig:3}}
  \end{figure}

The parameter contours for the difference between TE-only and EE-only parameter differences are shown in Figure 3. The expected value of the difference in parameter values is zero. The mean values in the contours are just the differences between the TE-only and EE-only means shown in Table 1. We use a Fisher forecast about these mean values with covariance matrix of the differences shown in Equation 13 to create the contours. Figure 3 illustrates that the parameter contours are within 1-2$\sigma$ of zero.

The SPTpol TE-only and EE-only parameters have a $\chi^2 = 4.16$ for five degrees of freedom (the parameter differences), which corresponds to a PTE of 0.53. This means that the SPTpol TE-only and EE-only parameter constraints are consistent. Without including the correlation between TE-only and EE-only parameters, the $\chi^2 = 3.09$ and has a PTE of 0.69, which is still consistent. \cite{Henning/etal:2018} report a PTE of 0.017 between the \lcdm\ predictions for the TE and EE power spectrum and their measured TE and EE power spectra. While this is only a mild tension of $2.1\sigma$, it is still different than the PTE values we find for the parameters. Since these are both consistency tests of SPTpol data that assume the \lcdm\ model, one might suspect that they should yield similar but not identical PTE values. We explore this below.

 To understand this difference in PTE values between consistency of the parameter and consistency of the power spectra with \lcdm\ based on these same parameters, we reexamine the 1000 simulated TE and EE band-powers used in Section 3.2. For each of the 1000 TE and EE simulated band-powers, there is a maximum likelihood parameter vector. We use the 10 parameter covariance matrix from Section 3.2 and Equation 13 to calculate the covariance matrix of the differences which is used to calculate 1000 $\chi^2$ and PTE values for the consistency of the parameters. This provides 1000 simulated versions of what we do for the real SPTpol TE-only and EE-only parameter vectors in this section. To simulate the consistency check with \lcdm\ that SPTpol performed, we compare the 1000 simulated band-powers to the maximum likelihood theory band-powers resulting from the maximum likelihood parameter finder. We calculate 1000 $\chi^2$ and PTE values for the simulated data. See Section 3.2 for more information about where each of these elements come from.  

We find that there is no correlation between the PTE for the consistency of the parameters and the PTE for the consistency of the simulated TE and EE spectra with \lcdm\ predictions. This statement holds so long as the assumptions made in the simulations are valid, specifically (1) \lcdm\ is the true model and (2) the SPTpol bandpower covariance matrix correctly describes the scatter in the TE and EE spectra. Based on the simulations, it is not surprising to find that the PTE from the parameters is qualitatively different from the PTE of the power spectra with \lcdm. This situation can arise because of statistical fluctuations. In general, evaluating consistency of both parameters and data is valuable. 

For SPTpol, the correlations between the TE-only and EE-only constraints on parameters are given by
\begin{align}
    \{H_0,\Omega_ch^2,\Omega_bh^2,\log(A_s),&n_s\}= \nonumber \\
   &\{9\%,10\%,8\%,25\%,32\%\}. \nonumber
\end{align}
These correlations are weak and positive. Because the correlations are weak, the effect of not including them in this case is small. The positive correlation between the TE-only and EE-only parameters is noteworthy because the TE and EE power spectra or data were negatively correlated. Both of these points can be seen in the relatively small shift in PTE from 0.53 to 0.69 when the correlations are not included. 

The positive correlations between TE-only and EE-only \lcdm\ parameters arise because the derivatives of the TE and EE theory spectra with respect to the \lcdm\ parameters happen to have predominantly opposite signs when weighted by the bandpower covariance matrices (see Equation 8). It is not always the case that the parameter correlations are positive. For example, we calculate the correlations between TE-only and EE-only \lcdm\ parameters for SPTpol when varying one parameter at a time and find that $\Omega_bh^2$ produces a negative correlation.

To investigate the impact of changing the assumed fiducial cosmology, we recalculate this matrix using the \planck\ TT, TE, EE + LowE cosmology given by \citep{planck/6:2018}
\begin{align}
    \{H_0,\Omega_ch^2,\Omega_bh^2,\log(A_s),&n_s\}= \nonumber \\
   &\{67.3,0.120,0.0224,3.05,0.963\}. \nonumber
\end{align}
For \planck\, we fix the value of $\tau = 0.054$. We find that using the \planck\ cosmology can shift the diagonal elements of the covariance matrix of the differences calculated using Equation 13 by as much as $25\%$. Nevertheless, when we calculate the level of correlations between SPTpol TE and EE parameters using the \planck\ cosmology as our fiducial model, we find that the change is only at the percent level:
\begin{align}
    \{H_0,\Omega_ch^2,\Omega_bh^2,\log(A_s),&n_s\}= \nonumber \\
   &\{9\%,10\%,9\%,26\%,33\%\}. \nonumber
\end{align}
We compare the contours from SPTpol shown in Figure 3 to contours from this \planck\ cosmology and find good agreement with an overall shrinking of contour size in the \planck\ fiducial model case. This implies that the structure of the covariance matrix remained the same, but the overall size of the covariance has been reduced. The $\chi^2$ and PTE using the \planck\ TT, TE, EE+LowE mean cosmology as the fiducial model when calculating the covariance matrix using Equation 7 are 6.2 and 0.29 respectively. Note that this is consistent with the overall reduction in the covariance. The important point is that changing the fiducial cosmology does not qualitatively alter the consistency of the TE-only and EE-only parameters even if it does shift the PTE value.

\section{The Cosmic Variance Limited Case}
In this section, we investigate the level and effect of correlation for the cosmic variance limited case to check if the results from Section 3 are specific to SPTpol as well as to project what future more precise CMB experiments can expect. We used \texttt{Pycamb} to generate TT, TE, and EE power spectra about a fiducial cosmology given by the joint fit in Table 1. We set $\tau$ to 0.078.  We find that using the \planck\ fiducial cosmology results in a negligible shift in the parameters indicating that the correlations are largely independent of fiducial cosmology within \lcdm. For the multipole covariance matrix, we use the cosmic variance limited multipole covariance matrix provided in Equation 7. To get the parameter covariance between parameters from correlated data sets, we use the analytic solution in Equation 7 where $\pmb{\mathbb{C}}^{XY}$ is the sub-block of the full covariance matrix. We calculate the derivative matrices about the fiducial cosmology using a simple finite difference method.

\subsection{Varying the Multipole Range}
We calculate the correlation matrix between all three combinations of TT, TE, and EE parameter constraints for variable multipole ranges. We fix the minimum multipole moment to be $\ell = 30$ and vary the maximum multipole moment between $\ell = 1000$ and $\ell = 4500$ representing varying noise levels truncating the effective maximum multipole moment. The lowest multipole moment is chosen to be $\ell = 30$ because this is the minimum mulipole moment where the likelihood is Gaussian. The results are shown in Figure 4. We then fix the maximum multipole moment to 2500 and vary the minimum multipole moment between 30 and 1000. The resulting plots of correlation coefficients are shown in Figure 5.

The correlations between the parameters from each of the various spectra combinations vary as a function of both maximum and minimum multipole moment included. In general the correlations for both TT-TE and TE-EE range from $\approx 0\%$ to $\approx 50\%$. The correlations between the TT-TE and TE-EE parameters are mostly positive regardless of the multipole range included. There are three exceptions to this that all involve TE-EE correlations of $\Omega_bh^2$ and are near $\ell = 1200$ in Figure 4 and near $\ell = 400$ and $\ell = 1000$ in Figure 5. The correlations between TT-EE are generally the weakest ranging from $-5\%$ to $12\%$. This is because there are weaker correlations between the TT and EE power spectra. Again there are a range of multipoles where the correlations between TE-EE for $\Omega_bh^2$ are less than the correlations TT-EE for $\Omega_bh^2$. There is an oscillatory structure in both Figures 4 and 5; however, there is not a regular period. This is a result of the TT, TE, and EE power spectra having peaks that are offset from one another. 

The weak correlations found for SPTpol in Section 3.3 are compatible with the correlations found in the cosmic variance limited case. This indicates that the level of correlations are not unique to SPTpol and in general weak correlations between parameters from different CMB power spectra should be expected. This poses an interesting question: if the correlations are weak, can we ignore them for the purposes of assessing consistency? In the next section we investigate the impact of not including these correlations on PTE values when making comparisons. 

\begin{figure*}[!tbp]
  \begin{minipage}[b]{0.4\textwidth}
    \hspace*{-0.4cm}
    \includegraphics[width=8in, height=8in]{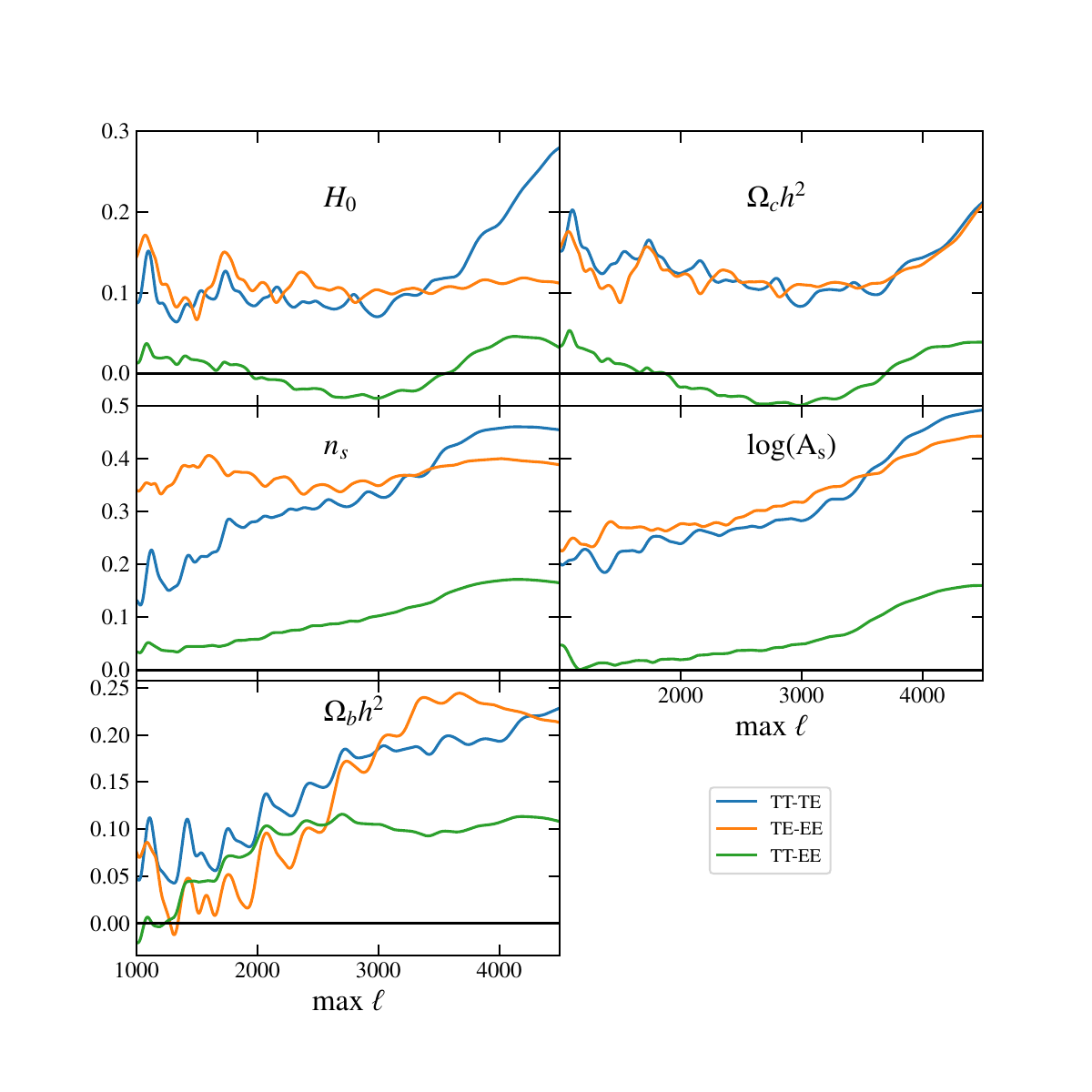}
  \end{minipage}
  \caption{The level of correlation between the parameters constraints for each of $H_0$, $\Omega_ch^2$, $\Omega_bh^2$, $\log(A_s)$, and $n_s$ for TT-only and EE-only (green), TT-only and TE-only (blue), TE-only and EE-only (orange) vary as a function of the maximum multipole moment included. The minimum multipole moment was fixed to $\ell = 30$. In general the correlation between TT-only and EE-only is the smallest. The correlations are compatible with the correlations found for SPTpol indicating that the SPTpol results are not unique. In all panels, there is a non-regular oscillatory structure that results from the TT, TE, and EE peaks being offset from one another. 
 \label{fig:4}}
  \end{figure*}

\begin{figure*}[!tbp]
  \begin{minipage}[b]{0.4\textwidth}
    \hspace*{-0.4cm}
    \includegraphics[width=8in, height=8in]{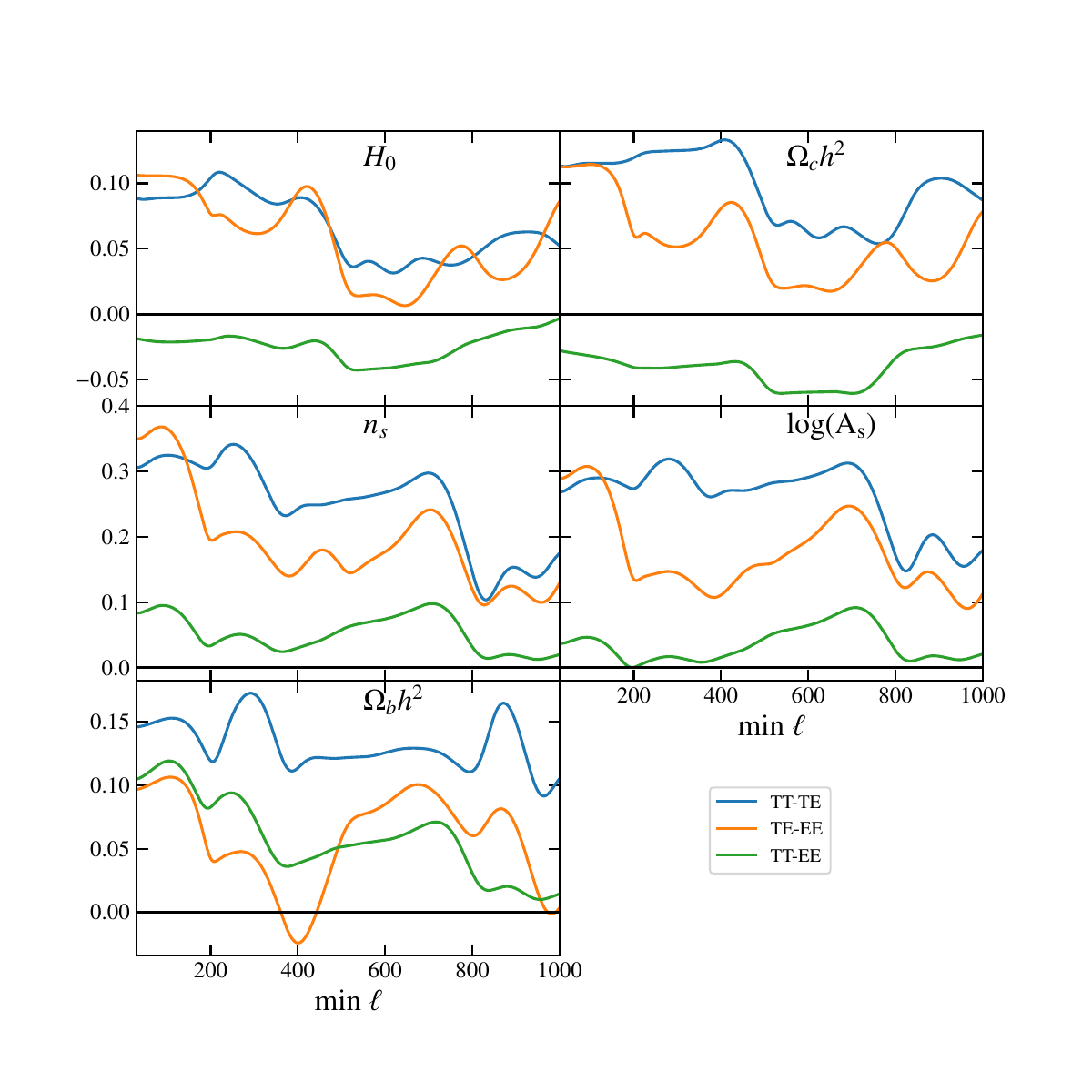}
  \end{minipage}
  \caption{The level of correlation between the parameters constraints for each of $H_0$, $\Omega_ch^2$, $\Omega_bh^2$, $\log(A_s)$, and $n_s$ for TT-only and EE-only (green), TT-only and TE-only (blue), TE-only and EE-only (orange) vary as a function of the minumum multipole moment included. We fix the maximum multipole moment to be $\ell = 2500$. In general, the correlations are weak ($< 40\%$), which is consistent with what we found for SPTpol. In all panels, there is a non-regular oscillatory structure that results from the offset in the TT, TE, and EE peaks.  
 \label{fig:5}}
  \end{figure*}

\begin{figure}[!tbp]
  \begin{minipage}[b]{0.4\textwidth}
    \hspace*{-0.4cm}
    \includegraphics[width=4in, height=4in]{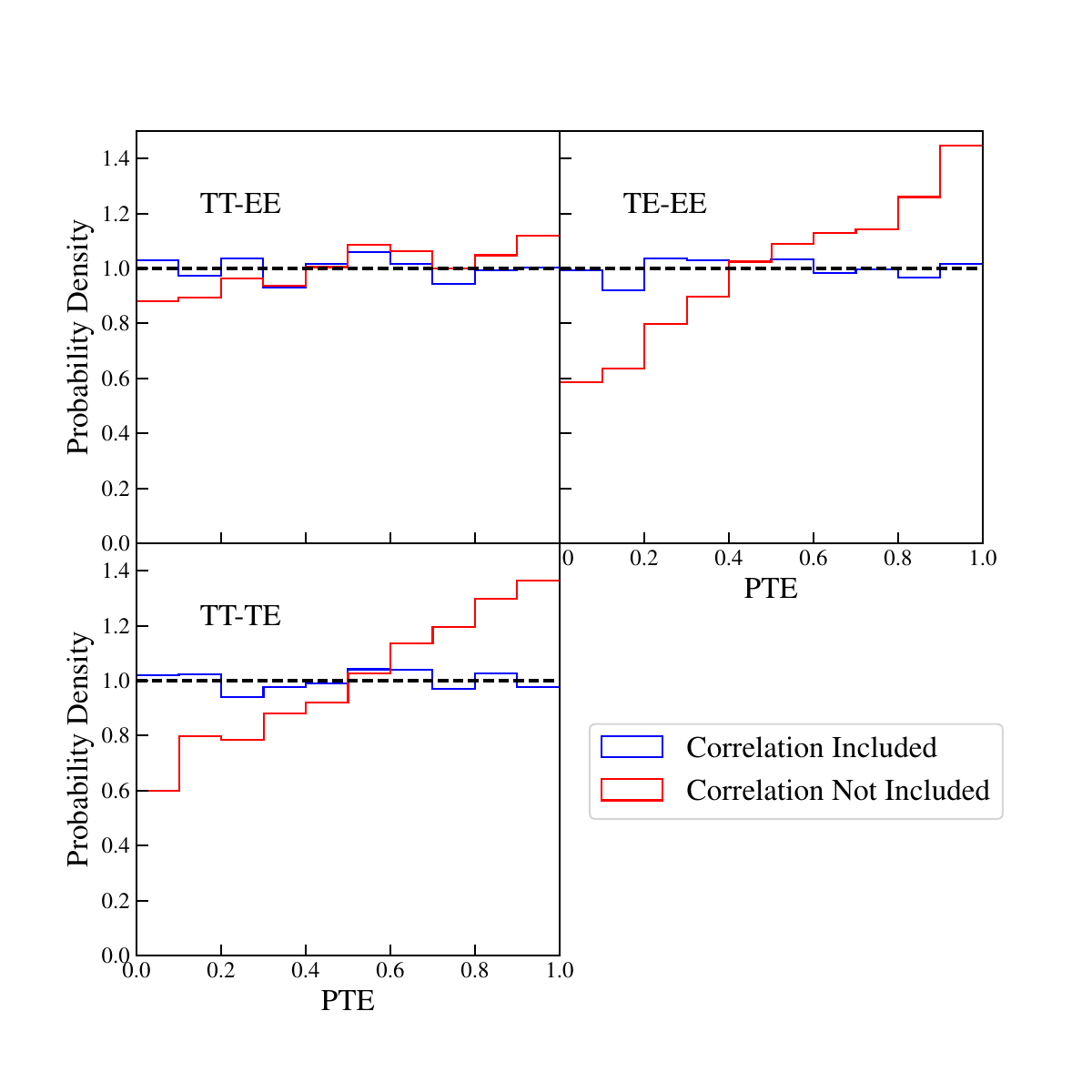}
  \end{minipage}
  \caption{Distribution of PTE values resulting from $\chi^2$ tests of the consistency of 10000 simulated parameter differences with zero. In blue, we take the correlation between the two spectra into account while in the red we do not. The black dashed line represents a uniform probability density function. There is a clear preference for TT-TE and TE-EE for larger PTE values or smaller $\chi^2$ values indicating that ignoring this parameter correlation will bias comparisons to look more consistent. There is a weaker trend for TT-EE resulting from a weaker correlation between the TT and EE power spectra. 
 \label{fig:6}}
  \end{figure}

\subsection{Impact of Neglecting Parameter Correlation in Consistency Checks}
We determine the effect of not including the correlation between the parameters from correlated data sub-sets on the eventual PTE value for consistency of parameter differences with zero. To do this, we utilize the relation between maximum likelihood and fiducial parameters found in Equation 6. Equation 6 shows that the maximum likelihood parameter vector $\bm{\theta}_{\textrm{ML}}$ can be calculated in terms of a set of fiducial parameters, $\bm{\theta}_\textrm{fid}$, and a data vector, $\bm{d}$. Again, we use the SPTpol joint fit as our fiducial cosmology. We calculate theory TT, TE, and EE power spectra over the multipole range $30\leq \ell \leq 2500$, and then use the cosmic variance limited multipole covariance matrix shown in Equation 9 to generate 10000 TT, TE, and EE power spectra from a normal distribution. Using these power spectra as the data vector $\bm{d}$ in Equation 6, we have 10000 TT, TE, and EE maximum likelihood parameter vectors. Note that this is equivalent to running the \texttt{CosmoMC} maximum likelihood finder for each of the 10000 TT, TE, and EE power spectra. 

We calculate two parameter covariance matrices for the cosmic variance limited case using Equation 7. In the first case, we include the correlation between the data encoded in the $\pmb{\mathbb{C}}$ sub-block of the multipole covariance matrix. In the second case, we set this correlation between the power spectra to zero. We calculate the $\chi^2$ and PTE values for each combination of power spectra for all 10000 realizations using both the parameter covariance matrix with correlations between the spectra and the parameter covariance matrix without correlations between the spectra. Histograms of the PTE values are shown in Figure 6. 

The probability distributions for the PTE values should be uniform, which is what we find when we include the correlations as evidenced by the good agreement between the blue histograms and the black dashed line in Figure 6. The PTE histograms for TE-EE and TT-TE show a strong preference for larger PTE values or smaller $\chi^2$ values in the case of no correlation. This follows because the correlations between these parameters are positive. Therefore, not including correlations will bias the $\chi^2$ to lower values or larger PTE values. Note that this is the worse of the two possibilities as making the parameters look more consistent than they are could result in an important systematic error being obscured. The TT-EE case also appears to have a trend preferring larger PTE values in the no correlation case, but it is much weaker. This is likely a result of the weaker correlation between the TT and EE power spectra. 

\section{Conclusions}

We derived an analytic expression for the covariance between model parameters from two correlated data sets. This analytic solution is an extension of the Fisher analysis with the added condition of a Gaussian likelihood. These assumptions can be applied to a wide variety of cosmological data sets. We showed using SPTpol simulations that the analytic solution returns results in good agreement with the maximum likelihood simulation method.  

We found that despite the fact that the TE and EE power spectra are negatively correlated, the \lcdm\ parameters from TE-only are positively correlated with the parameters from EE-only. We tested the consistency of the TE-only and EE-only parameters using the analytic solution and Equation 13 and found a PTE of 0.53 indicating the parameters are consistent. We further found that there is no correlation between consistency of the power spectra with \lcdm\ predictions and consistency of the \lcdm\ parameters resulting from the power spectra. For SPTpol, the correlations between the TE-only and EE-only determinations of $H_0$ is only 9$\%$; however, the correlations for log($A_s$) and $n_s$ are larger at 25$\%$ and  32$\%$.

We investigated the correlations between TT, TE, and EE power spectra as a function of the multipole range for the cosmic variance limited case. We found that the correlations between the TE-only and the TT-only or EE-only power spectra varied between $0\%$ and $50\%$ indicating weak levels of correlations. Of the \lcdm\ parameters, log($A_s$) and $n_s$ will have the largest correlations between TE-only and either TT-only and EE-only. The correlations between the TT-only and EE-only parameters are generally $<10\%$ in magnitude. This implies that future high resolution EE constraints will be largely independent of \planck\ TT constraints. Therefore, EE constraints will provide valuable new information that will constrain cosmological parameters and possible extensions to cosmological models \citep[e.g.,][]{Poulin/etal:2019,Lin/etal:2019}. We investigated the consequences of not including the correlation. Because the correlations are generally positive, the effect of not including the correlations is to erroneously make the parameters look more consistent than they are. We therefore recommend that these correlations be calculated and reported in future CMB analyses.

In this work, we quantified the consistency of sub-sets of CMB data, but this analytic solution has a wide variety of applications. For example, this method is applicable for different CMB experiments with sky overlap \citep[e.g.,][]{Louis/etal:2019}. It could also be applied to other cosmological measurements provided the assumptions listed in Section~2 hold, for example to check consistency across sub-sets of a supernova survey (e.g., different redshift bins) where systematic uncertainties are correlated across all the supernovae \citep[e.g.,][]{scolnic/etal:2018}.

The analytic solution complements simulations. Equation 7 provides a quick way to calculate what the expected level of correlation should be between parameters from correlated data sets. This analytic solution could be used in lieu of simulations or as a check that the simulations are working. Moreover, the analytic solution can be used as a check to gauge the impact of changing the model without the need to re-run potentially computationally intensive simulations.

\section{Acknowledgments}
This work was supported in part by NASA ROSES grants NNX16AF28G, NNX17AF34G, and 80NSSC19K0526. We acknowledge the use of the Legacy Archive for Microwave Background Data Analysis (LAMBDA), part of the High Energy Astrophysics Science Archive Center (HEASARC). HEASARC/LAMBDA is a service of the Astrophysics Science Division at the NASA Goddard Space Flight Center. This research project was conducted using computational resources at the Maryland Advanced Research Computing Center (MARCC). Figures 2 and 3 were created using the Python package GetDist. We thank 
Janet Weiland for helpful comments while completing the write up of this work. Finally, we would like to thank Yajing Huang for stimulating discussions.

\bibliography{cosmology,planck,act,spt}

\begin{thebibliography}{}
\expandafter\ifx\csname natexlab\endcsname\relax\def\natexlab#1{#1}\fi

\bibitem[{{Abbott} {et~al.}(2018){Abbott}, {Abdalla}, {Alarcon}, {Aleksi{\'c}},
  {Allam}, {Allen}, {Amara}, {Annis}, {Asorey}, {Avila}, {Bacon}, {Balbinot},
  {Banerji}, {Banik}, {Barkhouse}, {Baumer}, {Baxter}, {Bechtol}, {Becker},
  {Benoit-L{\'e}vy}, {Benson}, {Bernstein}, {Bertin}, {Blazek}, {Bridle},
  {Brooks}, {Brout}, {Buckley-Geer}, {Burke}, {Busha}, {Campos}, {Capozzi},
  {Carnero Rosell}, {Carrasco Kind}, {Carretero}, {Castander}, {Cawthon},
  {Chang}, {Chen}, {Childress}, {Choi}, {Conselice}, {Crittenden}, {Crocce},
  {Cunha}, {D'Andrea}, {da Costa}, {Das}, {Davis}, {Davis}, {De Vicente},
  {DePoy}, {DeRose}, {Desai}, {Diehl}, {Dietrich}, {Dodelson}, {Doel},
  {Drlica-Wagner}, {Eifler}, {Elliott}, {Elsner}, {Elvin-Poole}, {Estrada},
  {Evrard}, {Fang}, {Fernandez}, {Fert{\'e}}, {Finley}, {Flaugher}, {Fosalba},
  {Friedrich}, {Frieman}, {Garc{\'\i}a-Bellido}, {Garcia-Fernandez}, {Gatti},
  {Gaztanaga}, {Gerdes}, {Giannantonio}, {Gill}, {Glazebrook}, {Goldstein},
  {Gruen}, {Gruendl}, {Gschwend}, {Gutierrez}, {Hamilton}, {Hartley}, {Hinton},
  {Honscheid}, {Hoyle}, {Huterer}, {Jain}, {James}, {Jarvis}, {Jeltema},
  {Johnson}, {Johnson}, {Kacprzak}, {Kent}, {Kim}, {King}, {Kirk}, {Kokron},
  {Kovacs}, {Krause}, {Krawiec}, {Kremin}, {Kuehn}, {Kuhlmann}, {Kuropatkin},
  {Lacasa}, {Lahav}, {Li}, {Liddle}, {Lidman}, {Lima}, {Lin}, {MacCrann},
  {Maia}, {Makler}, {Manera}, {March}, {Marshall}, {Martini}, {McMahon},
  {Melchior}, {Menanteau}, {Miquel}, {Miranda}, {Mudd}, {Muir}, {M{\"o}ller},
  {Neilsen}, {Nichol}, {Nord}, {Nugent}, {Ogando}, {Palmese}, {Peacock},
  {Peiris}, {Peoples}, {Percival}, {Petravick}, {Plazas}, {Porredon}, {Prat},
  {Pujol}, {Rau}, {Refregier}, {Ricker}, {Roe}, {Rollins}, {Romer}, {Roodman},
  {Rosenfeld}, {Ross}, {Rozo}, {Rykoff}, {Sako}, {Salvador}, {Samuroff},
  {S{\'a}nchez}, {Sanchez}, {Santiago}, {Scarpine}, {Schindler}, {Scolnic},
  {Secco}, {Serrano}, {Sevilla-Noarbe}, {Sheldon}, {Smith}, {Smith}, {Smith},
  {Soares-Santos}, {Sobreira}, {Suchyta}, {Tarle}, {Thomas}, {Troxel},
  {Tucker}, {Tucker}, {Uddin}, {Varga}, {Vielzeuf}, {Vikram}, {Vivas},
  {Walker}, {Wang}, {Wechsler}, {Weller}, {Wester}, {Wolf}, {Yanny}, {Yuan},
  {Zenteno}, {Zhang}, {Zhang}, {Zuntz}, \& {Dark Energy Survey
  Collaboration}}]{Abbott/etal:2018}
{Abbott}, T.~M.~C., {Abdalla}, F.~B., {Alarcon}, A., {et~al.} 2018, \prd, 98,
  043526

\bibitem[{{Addison} {et~al.}(2016){Addison}, {Huang}, {Watts}, {Bennett},
  {Halpern}, {Hinshaw}, \& {Weiland}}]{addison/etal:2016}
{Addison}, G.~E., {Huang}, Y., {Watts}, D.~J., {et~al.} 2016, \apj, 818, 132

\bibitem[{{Addison} {et~al.}(2018){Addison}, {Watts}, {Bennett}, {Halpern},
  {Hinshaw}, \& {Weiland}}]{addison/etal:2018}
{Addison}, G.~E., {Watts}, D.~J., {Bennett}, C.~L., {et~al.} 2018, \apj, 853,
  119

\bibitem[{{Aylor} {et~al.}(2017){Aylor}, {Hou}, {Knox}, {Story}, {Benson},
  {Bleem}, {Carlstrom}, {Chang}, {Cho}, {Chown}, {Crawford}, {Crites}, {de
  Haan}, {Dobbs}, {Everett}, {George}, {Halverson}, {Harrington}, {Holder},
  {Holzapfel}, {Hrubes}, {Keisler}, {Lee}, {Leitch}, {Luong-Van}, {Marrone},
  {McMahon}, {Meyer}, {Millea}, {Mocanu}, {Mohr}, {Natoli}, {Omori}, {Padin},
  {Pryke}, {Reichardt}, {Ruhl}, {Sayre}, {Schaffer}, {Shirokoff},
  {Staniszewski}, {Stark}, {Vanderlinde}, {Vieira}, \&
  {Williamson}}]{Aylor/etal:2017}
{Aylor}, K., {Hou}, Z., {Knox}, L., {et~al.} 2017, \apj, 850, 101

\bibitem[{{Bennett} {et~al.}(2013){Bennett}, {Larson}, {Weiland}, {Jarosik},
  {Hinshaw}, {Odegard}, {Smith}, {Hill}, {Gold}, {Halpern}, {Komatsu}, {Nolta},
  {Page}, {Spergel}, {Wollack}, {Dunkley}, {Kogut}, {Limon}, {Meyer}, {Tucker},
  \& {Wright}}]{bennett/etal:2013}
{Bennett}, C.~L., {Larson}, D., {Weiland}, J.~L., {et~al.} 2013, \apjs, 208, 20

\bibitem[{{Benoit-L{\'e}vy} {et~al.}(2012){Benoit-L{\'e}vy}, {Smith}, \&
  {Hu}}]{benoit-levy/etal:2012}
{Benoit-L{\'e}vy}, A., {Smith}, K.~M., \& {Hu}, W. 2012, \prd, 86, 123008

\bibitem[{{Freedman} {et~al.}(2019){Freedman}, {Madore}, {Hatt}, {Hoyt},
  {Jang}, {Beaton}, {Burns}, {Lee}, {Monson}, {Neeley}, {Phillips}, {Rich}, \&
  {Seibert}}]{freedman/etal:2019}
{Freedman}, W.~L., {Madore}, B.~F., {Hatt}, D., {et~al.} 2019, \apj, 882, 34

\bibitem[{{Galli} {et~al.}(2014){Galli}, {Benabed}, {Bouchet}, {Cardoso},
  {Elsner}, {Hivon}, {Mangilli}, {Prunet}, \& {Wandelt}}]{Galli/etal:2014}
{Galli}, S., {Benabed}, K., {Bouchet}, F., {et~al.} 2014, \prd, 90, 063504

\bibitem[{Gelman {et~al.}(2013)Gelman, Carlin, Stern, Dunson, Vehtari, \&
  Rubin}]{gelman2013bayesian}
Gelman, A., Carlin, J., Stern, H., {et~al.} 2013, Bayesian Data Analysis, Third
  Edition, Chapman \& Hall/CRC Texts in Statistical Science (Taylor \& Francis)

\bibitem[{{Heavens}(2009)}]{Heavens:2009}
{Heavens}, A. 2009, ArXiv e-prints, arXiv:0906.0664

\bibitem[{{Henning} {et~al.}(2018){Henning}, {Sayre}, {Reichardt}, {Ade},
  {Anderson}, {Austermann}, {Beall}, {Bender}, {Benson}, {Bleem}, {Carlstrom},
  {Chang}, {Chiang}, {Cho}, {Citron}, {Corbett Moran}, {Crawford}, {Crites},
  {de Haan}, {Dobbs}, {Everett}, {Gallicchio}, {George}, {Gilbert},
  {Halverson}, {Harrington}, {Hilton}, {Holder}, {Holzapfel}, {Hoover}, {Hou},
  {Hrubes}, {Huang}, {Hubmayr}, {Irwin}, {Keisler}, {Knox}, {Lee}, {Leitch},
  {Li}, {Lowitz}, {Manzotti}, {McMahon}, {Meyer}, {Mocanu}, {Montgomery},
  {Nadolski}, {Natoli}, {Nibarger}, {Novosad}, {Padin}, {Pryke}, {Ruhl},
  {Saliwanchik}, {Schaffer}, {Sievers}, {Smecher}, {Stark}, {Story}, {Tucker},
  {Vanderlinde}, {Veach}, {Vieira}, {Wang}, {Whitehorn}, {Wu}, \&
  {Yefremenko}}]{Henning/etal:2018}
{Henning}, J.~W., {Sayre}, J.~T., {Reichardt}, C.~L., {et~al.} 2018, \apj, 852,
  97

\bibitem[{{Hikage} {et~al.}(2019){Hikage}, {Oguri}, {Hamana}, {More},
  {Mandelbaum}, {Takada}, {K{\"o}hlinger}, {Miyatake}, {Nishizawa}, {Aihara},
  {Armstrong}, {Bosch}, {Coupon}, {Ducout}, {Ho}, {Hsieh}, {Komiyama},
  {Lanusse}, {Leauthaud}, {Lupton}, {Medezinski}, {Mineo}, {Miyama},
  {Miyazaki}, {Murata}, {Murayama}, {Shirasaki}, {Sif{\'o}n}, {Simet},
  {Speagle}, {Spergel}, {Strauss}, {Sugiyama}, {Tanaka}, {Utsumi}, {Wang}, \&
  {Yamada}}]{Hikage/etal:2019}
{Hikage}, C., {Oguri}, M., {Hamana}, T., {et~al.} 2019, \pasj, 71, 43

\bibitem[{{Hildebrandt} {et~al.}(2018){Hildebrandt}, {K{\"o}hlinger}, {van den
  Busch}, {Joachimi}, {Heymans}, {Kannawadi}, {Wright}, {Asgari}, {Blake},
  {Hoekstra}, {Joudaki}, {Kuijken}, {Miller}, {Morrison}, {Tr{\"o}ster},
  {Amon}, {Archidiacono}, {Brieden}, {Choi}, {de Jong}, {Erben}, {Giblin},
  {Mead}, {Peacock}, {Radovich}, {Schneider}, {Sif{\'o}n}, \&
  {Tewes}}]{Hildebrandt/etal:2018}
{Hildebrandt}, H., {K{\"o}hlinger}, F., {van den Busch}, J.~L., {et~al.} 2018,
  arXiv e-prints, arXiv:1812.06076

\bibitem[{{Hou} {et~al.}(2018){Hou}, {Aylor}, {Benson}, {Bleem}, {Carlstrom},
  {Chang}, {Cho}, {Chown}, {Crawford}, {Crites}, {de Haan}, {Dobbs}, {Everett},
  {Follin}, {George}, {Halverson}, {Harrington}, {Holder}, {Holzapfel},
  {Hrubes}, {Keisler}, {Knox}, {Lee}, {Leitch}, {Luong-Van}, {Marrone},
  {McMahon}, {Meyer}, {Millea}, {Mocanu}, {Mohr}, {Natoli}, {Omori}, {Padin},
  {Pryke}, {Reichardt}, {Ruhl}, {Sayre}, {Schaffer}, {Shirokoff},
  {Staniszewski}, {Stark}, {Story}, {Vanderlinde}, {Vieira}, \&
  {Williamson}}]{Hou/etal:2018}
{Hou}, Z., {Aylor}, K., {Benson}, B.~A., {et~al.} 2018, \apj, 853, 3

\bibitem[{{Huang} {et~al.}(2019){Huang}, {Addison}, \&
  {Bennett}}]{Huang/etal:2019}
{Huang}, Y., {Addison}, G.~E., \& {Bennett}, C.~L. 2019, \apj, 882, 124

\bibitem[{{Huang} {et~al.}(2018){Huang}, {Addison}, {Weiland}, \&
  {Bennett}}]{Huang/etal:2018}
{Huang}, Y., {Addison}, G.~E., {Weiland}, J.~L., \& {Bennett}, C.~L. 2018,
  \apj, 869, 38

\bibitem[{{Joudaki} {et~al.}(2018){Joudaki}, {Blake}, {Johnson}, {Amon},
  {Asgari}, {Choi}, {Erben}, {Glazebrook}, {Harnois-D{\'e}raps}, {Heymans},
  {Hildebrand t}, {Hoekstra}, {Klaes}, {Kuijken}, {Lidman}, {Mead}, {Miller},
  {Parkinson}, {Poole}, {Schneider}, {Viola}, \& {Wolf}}]{Joudai/etal:2018}
{Joudaki}, S., {Blake}, C., {Johnson}, A., {et~al.} 2018, \mnras, 474, 4894

\bibitem[{{Kable} {et~al.}(2019){Kable}, {Addison}, \&
  {Bennett}}]{Kable/etal:2019}
{Kable}, J.~A., {Addison}, G.~E., \& {Bennett}, C.~L. 2019, \apj, 871, 77

\bibitem[{{Larson} {et~al.}(2015){Larson}, {Weiland}, {Hinshaw}, \&
  {Bennett}}]{larson/etal:2015}
{Larson}, D., {Weiland}, J.~L., {Hinshaw}, G., \& {Bennett}, C.~L. 2015, \apj,
  801, 9

\bibitem[{{Lewis} \& {Bridle}(2002)}]{lewis/bridle:2002}
{Lewis}, A., \& {Bridle}, S. 2002, \prd, 66, 103511

\bibitem[{{Lewis} {et~al.}(2000){Lewis}, {Challinor}, \&
  {Lasenby}}]{lewis/etal:2000}
{Lewis}, A., {Challinor}, A., \& {Lasenby}, A. 2000, \apj, 538, 473

\bibitem[{{Lin} {et~al.}(2019){Lin}, {Benevento}, {Hu}, \&
  {Raveri}}]{Lin/etal:2019}
{Lin}, M.-X., {Benevento}, G., {Hu}, W., \& {Raveri}, M. 2019, \prd, 100,
  063542

\bibitem[{{Lin} \& {Ishak}(2017)}]{Lin/etal:2017}
{Lin}, W., \& {Ishak}, M. 2017, \prd, 96, 083532

\bibitem[{{Louis} {et~al.}(2019){Louis}, {Garrido}, {Soussana}, {Tristram},
  {Henrot-Versill{\'e}}, \& {Vanneste}}]{Louis/etal:2019}
{Louis}, T., {Garrido}, X., {Soussana}, A., {et~al.} 2019, \prd, 100, 023518

\bibitem[{{Louis} {et~al.}(2014){Louis}, {Addison}, {Hasselfield}, {Bond},
  {Calabrese}, {Das}, {Devlin}, {Dunkley}, {D{\"u}nner}, {Gralla}, {Hajian},
  {Hincks}, {Hlozek}, {Huffenberger}, {Infante}, {Kosowsky}, {Marriage},
  {Moodley}, {N{\ae}ss}, {Niemack}, {Nolta}, {Page}, {Partridge}, {Sehgal},
  {Sievers}, {Spergel}, {Staggs}, {Walter}, \& {Wollack}}]{louis/etal:2014}
{Louis}, T., {Addison}, G.~E., {Hasselfield}, M., {et~al.} 2014, \jcap, 7, 16

\bibitem[{{Louis} {et~al.}(2017){Louis}, {Grace}, {Hasselfield}, {Lungu},
  {Maurin}, {Addison}, {Ade}, {Aiola}, {Allison}, {Amiri}, {Angile},
  {Battaglia}, {Beall}, {de Bernardis}, {Bond}, {Britton}, {Calabrese}, {Cho},
  {Choi}, {Coughlin}, {Crichton}, {Crowley}, {Datta}, {Devlin}, {Dicker},
  {Dunkley}, {D{\"u}nner}, {Ferraro}, {Fox}, {Gallardo}, {Gralla}, {Halpern},
  {Henderson}, {Hill}, {Hilton}, {Hilton}, {Hincks}, {Hlozek}, {Ho}, {Huang},
  {Hubmayr}, {Huffenberger}, {Hughes}, {Infante}, {Irwin}, {Muya Kasanda},
  {Klein}, {Koopman}, {Kosowsky}, {Li}, {Madhavacheril}, {Marriage}, {McMahon},
  {Menanteau}, {Moodley}, {Munson}, {Naess}, {Nati}, {Newburgh}, {Nibarger},
  {Niemack}, {Nolta}, {Nu{\~n}ez}, {Page}, {Pappas}, {Partridge}, {Rojas},
  {Schaan}, {Schmitt}, {Sehgal}, {Sherwin}, {Sievers}, {Simon}, {Spergel},
  {Staggs}, {Switzer}, {Thornton}, {Trac}, {Treu}, {Tucker}, {Van Engelen},
  {Ward}, \& {Wollack}}]{louis/etal:2017}
{Louis}, T., {Grace}, E., {Hasselfield}, M., {et~al.} 2017, \jcap, 6, 031

\bibitem[{{Manzotti} {et~al.}(2014){Manzotti}, {Hu}, \&
  {Benoit-L{\'e}vy}}]{Manzotti/etal:2014}
{Manzotti}, A., {Hu}, W., \& {Benoit-L{\'e}vy}, A. 2014, \prd, 90, 023003

\bibitem[{{McCarthy} {et~al.}(2018){McCarthy}, {Bird}, {Schaye},
  {Harnois-Deraps}, {Font}, \& {van Waerbeke}}]{McCarthy/etal:2018}
{McCarthy}, I.~G., {Bird}, S., {Schaye}, J., {et~al.} 2018, \mnras, 476, 2999

\bibitem[{{Mocanu} {et~al.}(2019){Mocanu}, {Crawford}, {Aylor}, {Benson},
  {Bleem}, {Carlstrom}, {Chang}, {Cho}, {Chown}, {Crites}, {de Haan}, {Dobbs},
  {Everett}, {George}, {Halverson}, {Harrington}, {Henning}, {Holder},
  {Holzapfel}, {Hou}, {Hrubes}, {Knox}, {Lee}, {Luong-Van}, {Marrone},
  {McMahon}, {Meyer}, {Millea}, {Mohr}, {Natoli}, {Omori}, {Padin}, {Pryke},
  {Reichardt}, {Ruhl}, {Sayre}, {Schaffer}, {Shirokoff}, {Staniszewski},
  {Stark}, {Story}, {Vanderlinde}, {Vieira}, {Williamson}, \&
  {Wu}}]{Mocanu/etal:2019}
{Mocanu}, L.~M., {Crawford}, T.~M., {Aylor}, K., {et~al.} 2019, \jcap, 2019,
  038

\bibitem[{{Motloch} \& {Hu}(2019)}]{Motloch/etal:2019}
{Motloch}, P., \& {Hu}, W. 2019, \prd, 99, 023506

\bibitem[{{Planck Collaboration} {et~al.}(2014){Planck Collaboration}, {Ade},
  {Aghanim}, {Armitage-Caplan}, {Arnaud}, {Ashdown}, {Atrio-Barand ela},
  {Aumont}, {Baccigalupi}, {Banday}, {Barreiro}, {Bartlett}, {Battaner},
  {Benabed}, {Beno{\^\i}t}, {Benoit-L{\'e}vy}, {Bernard}, {Bersanelli},
  {Bielewicz}, {Bobin}, {Bock}, {Bonaldi}, {Bond}, {Borrill}, {Bouchet},
  {Bridges}, {Bucher}, {Burigana}, {Butler}, {Calabrese}, {Cappellini},
  {Cardoso}, {Catalano}, {Challinor}, {Chamballu}, {Chary}, {Chen}, {Chiang},
  {Chiang}, {Christensen}, {Church}, {Clements}, {Colombi}, {Colombo},
  {Couchot}, {Coulais}, {Crill}, {Curto}, {Cuttaia}, {Danese}, {Davies},
  {Davis}, {de Bernardis}, {de Rosa}, {de Zotti}, {Delabrouille}, {Delouis},
  {D{\'e}sert}, {Dickinson}, {Diego}, {Dolag}, {Dole}, {Donzelli}, {Dor{\'e}},
  {Douspis}, {Dunkley}, {Dupac}, {Efstathiou}, {Elsner}, {En{\ss}lin},
  {Eriksen}, {Finelli}, {Forni}, {Frailis}, {Fraisse}, {Franceschi}, {Gaier},
  {Galeotta}, {Galli}, {Ganga}, {Giard}, {Giardino}, {Giraud-H{\'e}raud},
  {Gjerl{\o}w}, {Gonz{\'a}lez-Nuevo}, {G{\'o}rski}, {Gratton}, {Gregorio},
  {Gruppuso}, {Gudmundsson}, {Haissinski}, {Hamann}, {Hansen}, {Hanson},
  {Harrison}, {Henrot-Versill{\'e}}, {Hern{\'a}ndez-Monteagudo}, {Herranz},
  {Hildebrand t}, {Hivon}, {Hobson}, {Holmes}, {Hornstrup}, {Hou}, {Hovest},
  {Huffenberger}, {Jaffe}, {Jaffe}, {Jewell}, {Jones}, {Juvela},
  {Keih{\"a}nen}, {Keskitalo}, {Kisner}, {Kneissl}, {Knoche}, {Knox}, {Kunz},
  {Kurki-Suonio}, {Lagache}, {L{\"a}hteenm{\"a}ki}, {Lamarre}, {Lasenby},
  {Lattanzi}, {Laureijs}, {Lawrence}, {Leach}, {Leahy}, {Leonardi},
  {Le{\'o}n-Tavares}, {Lesgourgues}, {Lewis}, {Liguori}, {Lilje},
  {Linden-V{\o}rnle}, {L{\'o}pez-Caniego}, {Lubin}, {Mac{\'\i}as-P{\'e}rez},
  {Maffei}, {Maino}, {Mand olesi}, {Maris}, {Marshall}, {Martin},
  {Mart{\'\i}nez-Gonz{\'a}lez}, {Masi}, {Massardi}, {Matarrese}, {Matthai},
  {Mazzotta}, {Meinhold}, {Melchiorri}, {Melin}, {Mendes}, {Menegoni},
  {Mennella}, {Migliaccio}, {Millea}, {Mitra}, {Miville-Desch{\^e}nes},
  {Moneti}, {Montier}, {Morgante}, {Mortlock}, {Moss}, {Munshi}, {Murphy},
  {Naselsky}, {Nati}, {Natoli}, {Netterfield}, {N{\o}rgaard-Nielsen},
  {Noviello}, {Novikov}, {Novikov}, {O'Dwyer}, {Osborne}, {Oxborrow}, {Paci},
  {Pagano}, {Pajot}, {Paladini}, {Paoletti}, {Partridge}, {Pasian},
  {Patanchon}, {Pearson}, {Pearson}, {Peiris}, {Perdereau}, {Perotto},
  {Perrotta}, {Pettorino}, {Piacentini}, {Piat}, {Pierpaoli}, {Pietrobon},
  {Plaszczynski}, {Platania}, {Pointecouteau}, {Polenta}, {Ponthieu}, {Popa},
  {Poutanen}, {Pratt}, {Pr{\'e}zeau}, {Prunet}, {Puget}, {Rachen}, {Reach},
  {Rebolo}, {Reinecke}, {Remazeilles}, {Renault}, {Ricciardi}, {Riller},
  {Ristorcelli}, {Rocha}, {Rosset}, {Roudier}, {Rowan-Robinson},
  {Rubi{\~n}o-Mart{\'\i}n}, {Rusholme}, {Sandri}, {Santos}, {Savelainen},
  {Savini}, {Scott}, {Seiffert}, {Shellard}, {Spencer}, {Starck}, {Stolyarov},
  {Stompor}, {Sudiwala}, {Sunyaev}, {Sureau}, {Sutton}, {Suur-Uski}, {Sygnet},
  {Tauber}, {Tavagnacco}, {Terenzi}, {Toffolatti}, {Tomasi}, {Tristram},
  {Tucci}, {Tuovinen}, {T{\"u}rler}, {Umana}, {Valenziano}, {Valiviita}, {Van
  Tent}, {Vielva}, {Villa}, {Vittorio}, {Wade}, {Wandelt}, {Wehus}, {White},
  {White}, {Wilkinson}, {Yvon}, {Zacchei}, \& {Zonca}}]{planck/16:2013}
{Planck Collaboration}, {Ade}, P.~A.~R., {Aghanim}, N., {et~al.} 2014, \aap,
  571, A16

\bibitem[{{Planck Collaboration LI}(2017)}]{planck/51:2017}
{Planck Collaboration LI}. 2017, \aap, 607, A95

\bibitem[{{Planck Collaboration VI}(2018)}]{planck/6:2018}
{Planck Collaboration VI}. 2018, ArXiv e-prints, arXiv:1807.06209

\bibitem[{{Poulin} {et~al.}(2019){Poulin}, {Smith}, {Karwal}, \&
  {Kamionkowski}}]{Poulin/etal:2019}
{Poulin}, V., {Smith}, T.~L., {Karwal}, T., \& {Kamionkowski}, M. 2019, \prl,
  122, 221301

\bibitem[{{Raveri} \& {Hu}(2019)}]{Raveri/etal:2019}
{Raveri}, M., \& {Hu}, W. 2019, \prd, 99, 043506

\bibitem[{{Riess} {et~al.}(2019){Riess}, {Casertano}, {Yuan}, {Macri}, \&
  {Scolnic}}]{Riess/etal:2019}
{Riess}, A.~G., {Casertano}, S., {Yuan}, W., {Macri}, L.~M., \& {Scolnic}, D.
  2019, \apj, 876, 85

\bibitem[{{S{\'a}nchez} {et~al.}(2017){S{\'a}nchez}, {Grieb},
  {Salazar-Albornoz}, {Alam}, {Beutler}, {Ross}, {Brownstein}, {Chuang},
  {Cuesta}, {Eisenstein}, {Kitaura}, {Percival}, {Prada},
  {Rodr{\'\i}guez-Torres}, {Seo}, {Tinker}, {Tojeiro}, {Vargas-Maga{\~n}a},
  {Vazquez}, \& {Zhao}}]{Sanchez/etal:2017}
{S{\'a}nchez}, A.~G., {Grieb}, J.~N., {Salazar-Albornoz}, S., {et~al.} 2017,
  \mnras, 464, 1493

\bibitem[{{Scolnic} {et~al.}(2018){Scolnic}, {Jones}, {Rest}, {Pan},
  {Chornock}, {Foley}, {Huber}, {Kessler}, {Narayan}, {Riess}, {Rodney},
  {Berger}, {Brout}, {Challis}, {Drout}, {Finkbeiner}, {Lunnan}, {Kirshner},
  {Sanders}, {Schlafly}, {Smartt}, {Stubbs}, {Tonry}, {Wood-Vasey}, {Foley},
  {Hand}, {Johnson}, {Burgett}, {Chambers}, {Draper}, {Hodapp}, {Kaiser},
  {Kudritzki}, {Magnier}, {Metcalfe}, {Bresolin}, {Gall}, {Kotak}, {McCrum}, \&
  {Smith}}]{scolnic/etal:2018}
{Scolnic}, D.~M., {Jones}, D.~O., {Rest}, A., {et~al.} 2018, \apj, 859, 101

\bibitem[{{Scott} {et~al.}(1994){Scott}, {Srednicki}, \&
  {White}}]{scott/srednicki/white:1994}
{Scott}, D., {Srednicki}, M., \& {White}, M. 1994, \apjl, 421, L5

\bibitem[{{Sievers} {et~al.}(2013){Sievers}, {Hlozek}, {Nolta}, {Acquaviva},
  {Addison}, {Ade}, {Aguirre}, {Amiri}, {Appel}, {Barrientos}, {Battistelli},
  {Battaglia}, {Bond}, {Brown}, {Burger}, {Calabrese}, {Chervenak}, {Crichton},
  {Das}, {Devlin}, {Dicker}, {Bertrand Doriese}, {Dunkley}, {D{\"u}nner},
  {Essinger-Hileman}, {Faber}, {Fisher}, {Fowler}, {Gallardo}, {Gordon},
  {Gralla}, {Hajian}, {Halpern}, {Hasselfield}, {Hern{\'a}ndez-Monteagudo},
  {Hill}, {Hilton}, {Hilton}, {Hincks}, {Holtz}, {Huffenberger}, {Hughes},
  {Hughes}, {Infante}, {Irwin}, {Jacobson}, {Johnstone}, {Baptiste Juin},
  {Kaul}, {Klein}, {Kosowsky}, {Lau}, {Limon}, {Lin}, {Louis}, {Lupton},
  {Marriage}, {Marsden}, {Martocci}, {Mauskopf}, {McLaren}, {Menanteau},
  {Moodley}, {Moseley}, {Netterfield}, {Niemack}, {Page}, {Page}, {Parker},
  {Partridge}, {Plimpton}, {Quintana}, {Reese}, {Reid}, {Rojas}, {Sehgal},
  {Sherwin}, {Schmitt}, {Spergel}, {Staggs}, {Stryzak}, {Swetz}, {Switzer},
  {Thornton}, {Trac}, {Tucker}, {Uehara}, {Visnjic}, {Warne}, {Wilson},
  {Wollack}, {Zhao}, \& {Zunckel}}]{sievers/etal:2013}
{Sievers}, J.~L., {Hlozek}, R.~A., {Nolta}, M.~R., {et~al.} 2013, \jcap, 10, 60

\bibitem[{{Story} {et~al.}(2013){Story}, {Reichardt}, {Hou}, {Keisler}, {Aird},
  {Benson}, {Bleem}, {Carlstrom}, {Chang}, {Cho}, {Crawford}, {Crites}, {de
  Haan}, {Dobbs}, {Dudley}, {Follin}, {George}, {Halverson}, {Holder},
  {Holzapfel}, {Hoover}, {Hrubes}, {Joy}, {Knox}, {Lee}, {Leitch}, {Lueker},
  {Luong-Van}, {McMahon}, {Mehl}, {Meyer}, {Millea}, {Mohr}, {Montroy},
  {Padin}, {Plagge}, {Pryke}, {Ruhl}, {Sayre}, {Schaffer}, {Shaw}, {Shirokoff},
  {Spieler}, {Staniszewski}, {Stark}, {van Engelen}, {Vanderlinde}, {Vieira},
  {Williamson}, \& {Zahn}}]{story/etal:2013}
{Story}, K.~T., {Reichardt}, C.~L., {Hou}, Z., {et~al.} 2013, \apj, 779, 86

\bibitem[{{Verde}(2010)}]{Verde:2010}
{Verde}, L. 2010, in Lecture Notes in Physics, Berlin Springer Verlag, Vol.
  800, Lecture Notes in Physics, Berlin Springer Verlag, ed. G.~{Wolschin},
  147--177

\bibitem[{Weinberg(2008)}]{weinberg2008cosmology}
Weinberg, S. 2008, Cosmology, Cosmology (OUP Oxford)

\bibitem[{{Wong} {et~al.}(2019){Wong}, {Suyu}, {Chen}, {Rusu}, {Millon},
  {Sluse}, {Bonvin}, {Fassnacht}, {Taubenberger}, {Auger}, {Birrer}, {Chan},
  {Courbin}, {Hilbert}, {Tihhonova}, {Treu}, {Agnello}, {Ding}, {Jee},
  {Komatsu}, {Shajib}, {Sonnenfeld}, {Bland ford}, {Koopmans}, {Marshall}, \&
  {Meylan}}]{Holicow/etal:2019}
{Wong}, K.~C., {Suyu}, S.~H., {Chen}, G. C.~F., {et~al.} 2019, arXiv e-prints,
  arXiv:1907.04869

\bibitem[{{Yuan} {et~al.}(2019){Yuan}, {Riess}, {Macri}, {Casertano}, \&
  {Scolnic}}]{Yuan/etal:2019}
{Yuan}, W., {Riess}, A.~G., {Macri}, L.~M., {Casertano}, S., \& {Scolnic}, D.
  2019, arXiv e-prints, arXiv:1908.00993

\end{thebibliography}
\end{document}